\documentstyle[11pt]{article}

\input epsf

\begin{document}

{\sf
\hspace*{\fill}SU-4240-682\\
\hspace*{\fill}August 1998\\
}
\vspace*{8mm}

\begin{center}
{\LARGE \bf Phase diagram of three-dimensional dynamical triangulations
 with a boundary}
 
Simeon Warner$^*$, Simon Catterall$^*$, Ray Renken$^{\dagger}$\\
$^*$ Department of Physics, Syracuse University, 
  Syracuse, NY 13210, USA\\
$^{\dagger}$ Department of Physics, Unviersity of Central Florida,
  Orlando, FL 32816, USA
\end{center}

%%\footnotetext{Corresponding author: Simeon Warner, Department of Physics, Syracuse University, 
%%               Syracuse, NY 13210, USA. Tel: (+1) (315) 443 3985, fax: (+1) (315) 443 9103, email: {\tt simeon@physics.syr.edu}}

\begin{abstract}
We use Monte Carlo simulation to study the phase diagram of 
three-dimensional dynamical triangulations with a boundary. 
Three phases are indentified and characterized. 
One of these phases is a new, boundary dominated phase; 
a simple argument is presented to explain its existence. 
First-order transitions are shown to occur along the critical 
lines separating phases.
\end{abstract}

\section*{Dynamical triangulations with a boundary term}
  
Dynamical triangulation models arise from
simplicial discretizations of continuous Riemannnian manifolds.
A manifold is approximated by glueing together a set of
equilateral simplices with fixed edgelengths. This glueing
ensures that each face is shared by exactly two distinct simplices --
the resultant simplicial lattice is called a triangulation.
In the context of Euclidean quantum gravity it is
natural to consider a weighted sum of all possible triangulations as a 
candidate for a regularized path integral over metrics. Physically
distinct metrics correspond to inequivalent simplicial
triangulations. This prescription
has been shown to be very successful in two-dimensions 
(see, for example,~\cite{AMBJORN94}). 

Most analytic studies and almost all numerical work done so far has 
been restricted to compact manifolds like the sphere. In this
paper we develop techniques that allow us to extend numerical
studies to simplicial manifolds with boundaries. Specifically, we
study the $3$-disk created by inserting an $S^2$ boundary into
a triangulation of the sphere $S^3$. This allows us to
compute an object which is the simplicial equivalent of the
`wavefunction of the Universe' \cite{HARTLE+83}:
\begin{equation}
\psi\left[h\right]=\int Dg e^{-S\left(g\right)}
\end{equation}
The functional integral over 3-metrics $g$ is restricted to those
with 2-metric $h$ on the boundary. The simplicial analog is
simply
\begin{equation}
\psi\left(T_2\right)=\sum_{T_3} e^{-S_L\left(T_3\right)}
\end{equation}
Thus the probability amplitude for finding a particular 2-triangulation
$T_2$ is obtained by counting (with some weight) all
3-triangulations $T_3$ which contain $T_2$ as their boundary.
A natural lattice action $S_L$ can be derived from the continuum action
by straightforward techniques \cite{HARTLE+81}. It contains
both the usual Regge curvature piece familiar from
compact triangulations together with a boundary term.
The boundary term arises from discretization of the extrinsic
curvature of the boundary embedded in the bulk.
In three-dimensions the curvature 
is localized on links. If $L_M$ denotes the set of links in the
bulk of the 3-triangulation (excluding the boundary) and $L_{\partial M}$
those in the boundary the action can be written 
\begin{equation}
  S_{L} = \kappa_1\left(\sum_{h\in L_M}\left(2\pi-\alpha n_h\right)+
                        \sum_{h\in L_{\partial M}}\left(\pi-\alpha n_h\right)\right)						
\end{equation} 
The quantity $\alpha=\arccos{(1/3)}$ and $n_h$ is the number of simplices
sharing the link (hinge) $h$. Typically $S_L$ will also contain
a bulk cosmological constant that can be used to tune the
simulation volume. The resultant action can be rewritten in the
form 
\begin{equation}
  S_{b} = - \kappa_0 N_0 + \kappa_3 N_3 + \kappa_b N_2^b \label{3daction}
\end{equation}
where $N_2^b$ is the area of the boundary. 
Here, $\kappa_3$ is used to tune the volume of the system. We are thus left
with a two-dimensional phase space parameterized by $\kappa_0$ 
and $\kappa_b$ conjugate to the number of vertices and the number of
boundary triangles.
It is trivial to generalize this to, for example, four-dimensions.
The partition function for the system is then
\begin{equation}
  Z = \sum_{T} e^{-S_{b}}
\end{equation}
where the sum is over triangulations, $T$. 

Various other extended phase diagrams have been studied for
three-dimensional dynamical triangulations including 
adding spin matter \cite{RENKEN+94,AMBJORN+93}, 
adding gauge matter \cite{RENKEN+93,AMBJORN+94}, and 
adding a measure term \cite{RENKEN+97}. 
Much of this work was motivated by the desire to
find a continuous phase transition. No such transitions have been found.

\section*{Simulation}

Our simulation algorithm is an extension of the algorithm for 
compact manifolds in arbitrary dimension described by 
Catterall~\cite{CATTERALL95}. Consider the environment of
any vertex in a $D$-triangulation --- 
it is composed of simplices making up
a trivial $D$-ball. The boundary of this $D$-ball is just
the sphere $S^{(D-1)}$. A boundary with the topology of $S^{(D-1)}$ can thus be created
in the original triangulation by removing these simplices. If the
original triangulation corresponded to the sphere $S^D$ the 
topology of the new triangulation is that of a $D$-disk.

In practice we simulate a compact manifold with one `special'
vertex. This vertex and all the simplices sharing it are ignored during
any measurement. In this way every triangulation of
our marked sphere $S^D$ is in one-to-one correspondence with
a triangulation of the $D$-disk. Notice that the usual compact manifold
moves applied to all simplices (including those sharing the
special vertex) will in general change the boundary of the $D$-disk. 
Indeed these moves are ergodic with respect to the boundary.
Furthermore, the proof that these moves satisfy a detailed balance
relation goes through just as for the compact case. The one
extra restriction is simple --- one must never delete the
special vertex. With this trick we
can trivially extend our compact codes to the situation in
which a $S^{(D-1)}$ boundary has been added. We are merely simulating
a compact lattice with an action which singles out a special vertex
and its neighbour simplices. This contrasts with the set of 
additional boundary moves used by Adi {\em et al}~\cite{ADI+93}
for simulations in two-dimensions.

Measurements do not include the special vertex or
any simplices connected to it. For example, the number of $D$-simplices
in the system with boundary is the number of $D$-simplices in the
whole simulation minus the number of $D$-simplices sharing the special 
vertex. The size of the boundary is simply 
the number of $D$-simplices sharing the special vertex.

We have used the Metropolis Monte Carlo~\cite{METROPOLIS+53} 
scheme with usual update rule:
\begin{equation}
  p({\rm accept\ move}) = min \{ e^{-\Delta S_{b}} , 1 \}
\end{equation}
and in this way we explore the space of unlabeled triangulations
with the action $S_{b}$ (equation~\ref{3daction} for three-dimensions).

\subsection*{Checks in two-dimensions}

In two dimensions we tested our simulation code at small volumes by comparing with
hand calculated amplitudes for small disks. We label disk configurations
by the number of triangles and the boundary length: $(N_2,N_1^b)$.
We calculated the ratios of amplitudes for disks (1,1):(2,4):(3,3):(3,5)
to be 1:1.5:1:3. Our simulation gave 1.01:1.52:1:3.04 from a sample
of 1 million disks with volume 1--3. This test was extended up to
volume 5 disks, also showing good agreement.

We also tested our simulation code by comparing results for two-dimensions
of Adi {\em et al}~\cite{ADI+93}. All our results agree within
the statistical errors. Table~\ref{adi_table} shows a comparison of the
results for a selection of lattice sizes.

\begin{table}[ht]
\begin{center}
\begin{tabular}{|r|r|r|}
\hline
$N_2$	&	$\langle N_1^b \rangle$ (Adi {\em et al})	
        &	$\langle N_1^b \rangle$ (this work) \\
\hline
50	&	39.83(5)	&	39.88(4) 	\\
100	&	78.05(5)	&	78.09(4) 	\\
200	&	154.54(6)	&	154.57(6)	\\
400	&	307.62(8)	&	307.53(8)	\\
800	&	613.4(1)	&	613.4(1)	\\
1600	&	1225.1(1)	&	1225.1(2)	\\
3200	&	2448.7(2)	&	2448.8(3)	\\
6400	&	4895.6(3)	&	4895.4(13)	\\
\hline
\end{tabular}
\end{center}
\caption{Estimates of $\langle N_1^b \rangle$, the expectation
value of the boundary size (length), for two-dimensional manifolds of
various sizes from Adi {\em et al}~\protect\cite{ADI+93}, and from this work.}
\label{adi_table}
\end{table}

\section*{Phase diagram}

We performed a set of simulations in three-dimensions with 
action of equation~\ref{3daction}. In all runs 
$\kappa_3$ was used to tune the nominal system volume, $N_3$, to
2000 for each given $\kappa_0$ and $\kappa_b$.

\begin{figure}[htp]
\epsfysize=78mm
\epsfbox{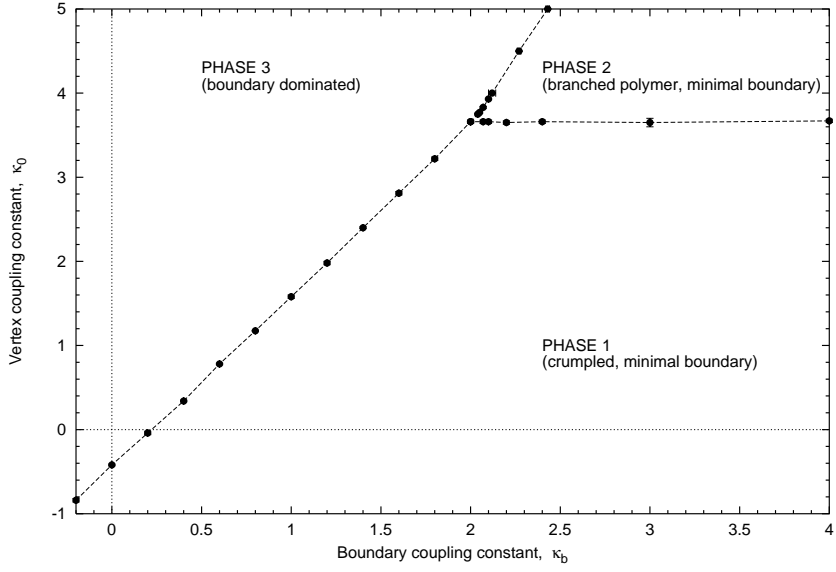}
\caption{\label{phaseplot} Phase diagram for $3$-dimensional dynamical
triangulation with a boundary. 
All points have error bars in either $\kappa_b$ or $\kappa_0$, 
most cannot be seen because they are smaller than the symbols.
Nominal simulation volume, $N_3 = 2000$.}
\end{figure}

In three-dimensions there are just 4 types of move: vertex insertion,
vertex deletion and exchange of a link with a face (two moves: 
link to face or face to link). Where these moves take place on 
sections of the triangulation involving the special vertex we take 
care to count changes in the numbers of simplices inside and 
outside of the boundary but otherwise the moves are the same as 
for the bulk.
Series of runs varying either $\kappa_0$ or $\kappa_b$ were made
and the vertex susceptibility used to search for phase transitions.
We define the vertex susceptibility, $\chi$, to be normalized with
respect to the number of 3-simplices:
\begin{equation}
  \chi = \frac{1}{N_3} ( \langle N_0^2\rangle - \langle N_0\rangle^2 ) 
\end{equation}
The points shown in figure~\ref{phaseplot} are taken from the positions
of peaks in the vertex susceptibility.

\begin{figure}[htp]
\epsfysize=78mm
\epsfbox{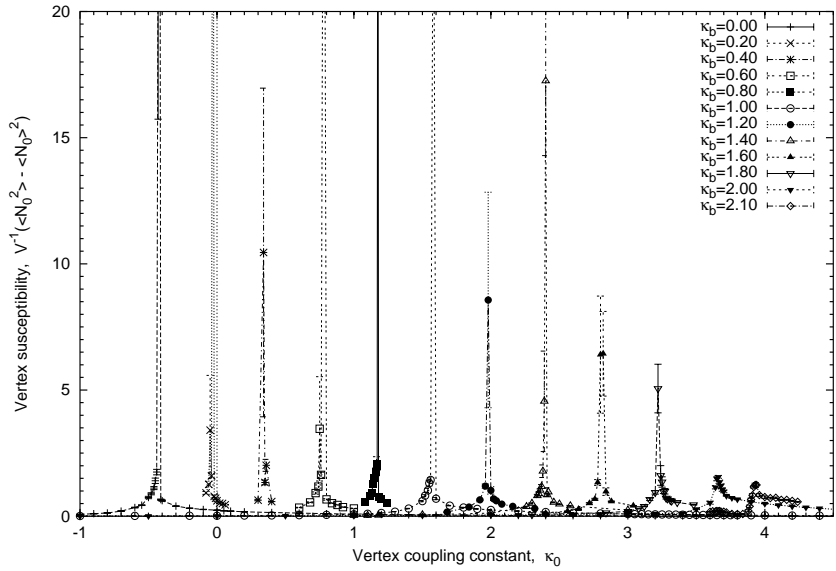}
\caption{\label{suscplot} Sample of vertex susceptibility data for different
values of the boundary coupling constant,~$\kappa_b$.}
\end{figure} 

In figure~\ref{phaseplot} there are three phases which we characterize as:
phase 1 - crumpled, minimal boundary; 
phase 2 - branched-polymer, minimal boundary; and
phase 3 - boundary dominated.
In phases 1 and 2 the boundary is simply 4 triangles (2-simplices) connected
to form a tetrahedral hole. The system is essentially like a compact
manifold with one marked 3-simplex --- the tetrahedral hole.
In phase 3 the boundary is large --- typically a substantial fraction of
the bulk volume.

\clearpage

\begin{figure}[ht]
\epsfbox{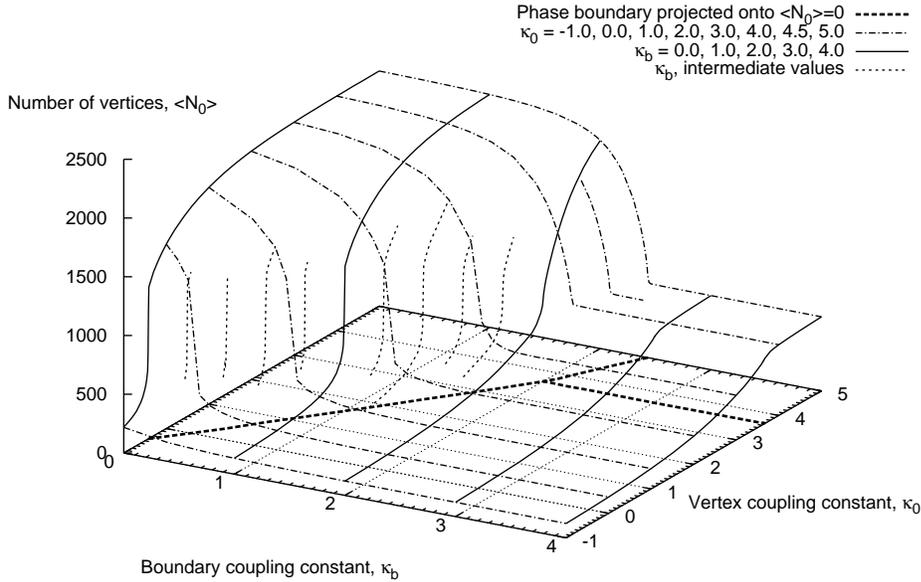}
\caption{Number of vertces, $\langle N_0 \rangle$, as a function of $\kappa_0$ and $\kappa_b$.
Nominal simulation volume, $N_3 = 2000$. Note that we see three distinct
areas with different values of $\langle N_0 \rangle$: the boundary dominated phase (small 
$\kappa_b$, large $\kappa_0$) with large $\langle N_0 \rangle$, the crumpled 
phase (small $\kappa_0$) with small $\langle N_0 \rangle$, and the branched-polymer phase
(large $\kappa_b$ and $\kappa_0$) with intermediate $\langle N_0 \rangle$.
}
\end{figure} 

\subsection*{Simple argument for boundary dominated phase}

Here we argue that the boundary dominated phase can be explained
by considering an effective action written in terms of the
boundary size. We show that in certain circumstances a large
boundary will decrease this action. Otherwise one of the minimal
boundary phases will be favored.

Consider the action:
\begin{equation}
  S_{b} = -k_0 N_0 + k_b N_2^b
\end{equation}
We ignore the volume term as this is kept fixed during the simulation. If
we note that the boundary is itself a 2-sphere then we know that:
\begin{equation} 
  N_2^b = 2 (N_0^b -2) 
\end{equation}
and 
\begin{equation} 
  N_0 = N_0^b + N_0^i
\end{equation}
where 
$N_0$ is the number of vertices, 
$N_0^b$ is the number of vertices on the boundary, 
$N_0^i$ is the number of internal vertices,  and
$N_2^b$ is the number of 2-simplices (triangles) on the boundary. 
We may thus rewrite the action:
\begin{equation}
  S_{b} = -\kappa_0 N_0^i + (2\kappa_b - \kappa_0 ) N_0^b
\end{equation}
If we now consider $N_0^i$ fixed and note that the number of manifolds with
boundary size $N_2^b$ is governed by an exponential factor 
$e^{\kappa_b^c N_2^b} = e^{2 \kappa_b^c N_0^b}$, where $\kappa_b^c$ is
a new constant, we may then write an
effective action for the number of boundary vertices:
\begin{equation}
  S_{eff} \approx (-\kappa_0 + 2(\kappa_b - \kappa_b^c)) N_0^b
\end{equation}
The presence of small or large boundaries is then
determined by the sign of this action. We thus expect the phase transition
at $\kappa_0 = 2 (\kappa_b - \kappa_b^c)$ which is in good agreement with what 
we see.

\subsection*{Order of transitions}

\begin{figure}[ht]
\epsfbox{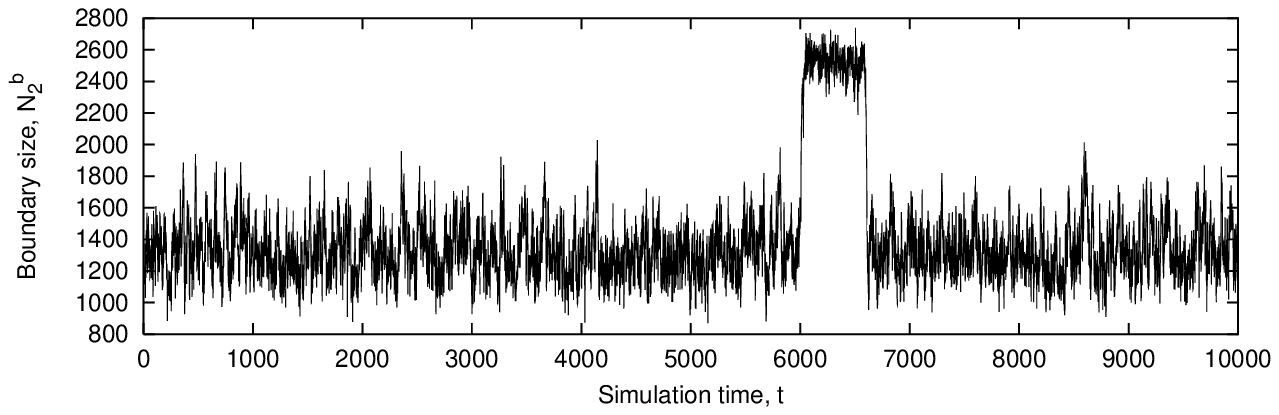}
\epsfbox{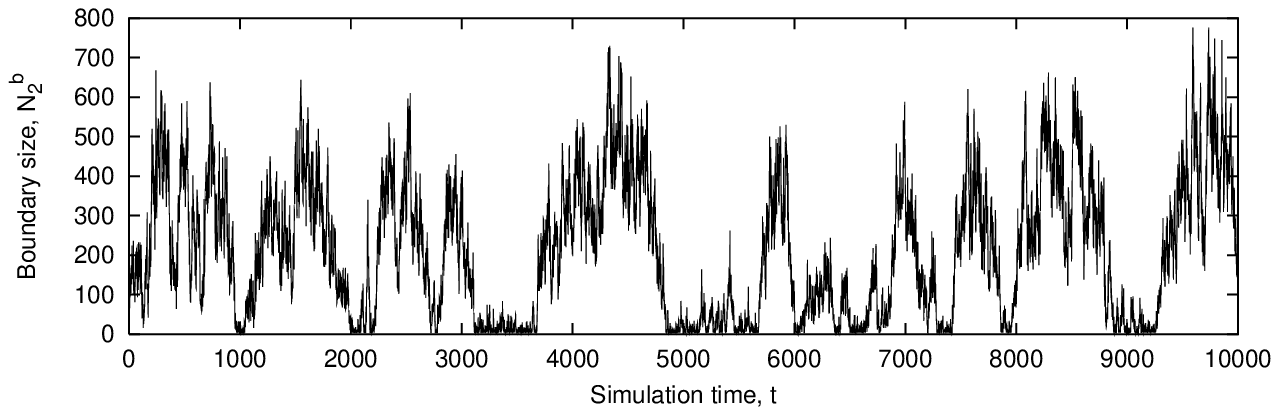}
\caption{\label{bistabplot} Time series showing the boundary size ($N_2^b$)
 during simulation. The upper plot is at the transition between the crumpled
 and boundary dominated phases ($k_0 = -0.423$, $k_b = 0$). The lower plot is
 at the transition between the branched-polymer and boundary dominated phases 
 ($k_0 = 5$, $k_b = 2.43$). Nominal simulation volume, $N_3 = 2000$, and time is in 
 units of $100N_3$ attempted updates.}
\end{figure} 

Simulations of compact manifolds in three and four-dimensions are known to 
have a first-order phase transition between crumpled and branched 
polymer phases~(3d~\cite{AMBJORN+92b}, 4d~\cite{BIALAS+96,DEBAKKER96}).
Our Monte Carlo time series show strong bistability on all three
phase boundaries (see figure~\ref{bistabplot}). We take this to 
indicate that all three phase transitions are first-order.

\section*{Concluding remarks}

We have demonstrated an arbitrary dimension algorithm for simulating
dynamical triangulations with a boundary. This has been tested against known
results in two-dimensions and used to map the phase diagram in 
three-dimensions.

We have identified three phases in three-dimensional dynamical triangulations 
with a boundary and mapped the boundaries within the range of couplings 
$-1 < \kappa_0 < 5$ and $-0.5 < \kappa_b < 4$.
The observed phases include the crumpled and branched-polymer phases
seen in triangulations of compact manifolds, and also a new, boundary
dominated phase. The existence of this phase, and the shape of the phase
boundary on the $\kappa_0$--$\kappa_b$ phase diagram, is predicted by a 
simple argument. Obvious bistability in the time series at the
phase transitions indicates that all transitions within the range 
of couplings studied are first-order.

\section*{Acknowledgments}

Simon Catterall was supported in part by DOE grant DE-FG02-85ER40237. 
Ray Renken was supported in part by NSF grant PHY-9503371.

%%%%%%%%%%%%%%%%%%%%%%%%%%%%%%%%%%%%%%%%%%%%%%%%%%%%%%%%%%%%%%%%%%%%%%%%%%


\begin{thebibliography}{10}

\bibitem{ADI+93}
E~Adi, M~Hasenbusch, M~Marcu, E~Pazy, K~Pinn and S~Solomon.
\newblock Monte carlo simulation of 2-d quantum gravity as open dynamically
  triangulated random surfaces.
\newblock {\em XXX archive} {\tt hep-lat/9310016} (1993).

\bibitem{AMBJORN94}
J~Ambj\/{o}rn.
\newblock Quantization of geometry.
\newblock {\em XXX archive} {\tt hep-th/9411179} (1994).

\bibitem{AMBJORN+92b}
J~Ambj\/{o}rn, D~V Boulatov, A~Krzywicki and S~Varsted.
\newblock The vacuum in three-dimensaional simplicial quantum gravity.
\newblock {\em Physiscs Letters B}, {\bf 276} 432--436 (1992).

\bibitem{AMBJORN+94}
J~Ambj\/{o}rn, J~Jurkiewicz, S~Bilke, Z~Burda and B~Petersson.
\newblock $z_2$ gauge matter coupled to 4-d simplicial quantum gravity.
\newblock {\em Modern Physics Letters A}, {\bf 9} 2527--2541 (1994).

\bibitem{AMBJORN+93}
J~Ambj\/{o}rn, C~Kristjansen, Z~Burda and J~Jurkiewicz.
\newblock {\em Nuclear Physics B Proc. Suppl.}, 771-- (1993).

\bibitem{BIALAS+96}
P~Bialas, Z~Burda, A~Krzywicki and B~Petersson.
\newblock Focusing on the fixed point of 4d simplicial gravity.
\newblock {\em Nuclear Physics B}, {\bf 472} 293--308 (1996).
\newblock Also {\em XXX archive} {\tt hep-lat/9601024}.

\bibitem{CATTERALL95}
S~Catterall.
\newblock Simulations of dynamically triangulated gravity - an algoritm for
  arbitrary dimension.
\newblock {\em Computer Physics Communications}, {\bf 87} 409--415 (1995).

\bibitem{DEBAKKER96}
B~{de Bakker}.
\newblock Further evidence that the transition of 4d dynamical triangulation is
  1$^st$ order.
\newblock {\em Physiscs Letters B}, {\bf 389} 238--242 (1996).
\newblock Also {\em XXX archive} {\tt hep-lat/9603024}.

\bibitem{HARTLE+83}
J~B Hartle and S~W Hawking.
\newblock Wave function of the universe.
\newblock {\em Physical Review D}, {\bf 12} 2960--2975 (1983).

\bibitem{HARTLE+81}
J~B Hartle and R~Sorkin.
\newblock Boundary terms in the action for the regge calculus.
\newblock {\em General Relativity and Gravity}, {\bf 13} 541--549 (1981).

\bibitem{METROPOLIS+53}
Nicholas Metropolis, Arianna~W Rosenbluth, Marshall~N Rosenbluth, Augusta~H
  Teller and Edward Teller.
\newblock Equation of state calculations by fast computing machines.
\newblock {\em The Journal of Chemical Physics}, {\bf 21} (1953).

\bibitem{RENKEN+93}
Ray~L Renken, Simon~M Catterall and John~B Kogut.
\newblock Three dimensional quantum gravity coupled to ising matter.
\newblock {\em Nuclear Physics B}, {\bf 389} 601--610 (1993).

\bibitem{RENKEN+94}
Ray~L Renken, Simon~M Catterall and John~B Kogut.
\newblock Three-dimensional quantum gravity coupled to gauge fields.
\newblock {\em Nuclear Physics B}, {\bf 422} 677--689 (1994).

\bibitem{RENKEN+97}
Ray~L Renken, Simon~M Catterall and John~B Kogut.
\newblock Phase structure of dynamical triangulation models in three
  dimensions.
\newblock {\em XXX archive {\tt hep-lat/9712011}} (1997).

\end{thebibliography}
\end{document}